\documentclass{article}

\usepackage{arxiv}

\usepackage[utf8]{inputenc} 
\usepackage[T1]{fontenc}    
\usepackage{hyperref}       
\usepackage{url}            
\usepackage{booktabs}       
\usepackage{nicefrac}       
\usepackage{microtype}      
\usepackage{lipsum}		

\usepackage{dblfloatfix}
\usepackage{cite}
\usepackage{amsmath,amssymb,amsfonts}
\usepackage{algorithmic}
\usepackage{graphicx}
\usepackage{textcomp}
\usepackage{caption,subcaption}
\usepackage{array}
\usepackage{epsfig}
\usepackage{epstopdf}
\usepackage{multicol}
\usepackage[ruled,vlined,linesnumbered,noresetcount]{algorithm2e}
\usepackage{mathtools}
\usepackage{cuted}
\usepackage{blindtext}
\usepackage{url}
\usepackage{soul}
\usepackage{color}
\usepackage{longtable}
\usepackage{enumitem}
\usepackage{multirow}

\title{A Game-Theoretic Approach for Enhancing Security and Data Trustworthiness in IoT Applications}

\author{
  Mohamed S. Abdalzaher$^{1,2}$\thanks{$^2$National Research Institute of Astronomy and Geophysics, Seismology Department, 11421, Egypt} $\;$ and Osamu Muta$^1$ \\
  $^1$Center for Japan-Egypt Cooperation\\ in Science and Technology, \\Kyushu University, \\744 Motooka, Nishi-ku, Fukuoka-shi 819-0395, Japan
  \texttt{mohamed.abdelzaher@ejust.edu.eg} \\ }
  
\begin{document}
\maketitle

\begin{abstract}
Wireless sensor networks (WSNs)-based internet of things (IoT)  are among the fast booming technologies that drastically contribute to different systems management and resilience data accessibility. Designing a robust IoT network imposes some challenges such as data trustworthiness (DT) and power management. This paper presents a repeated game model to enhance clustered WSNs-based IoT security and DT against the selective forwarding (SF) attack. Besides, the model is capable of detecting the hardware (HW) failure of the cluster members (CMs) and conserve the power consumption due to packet retransmission. The model relies on TDMA protocol to facilitate the detection process and to avoid collision between the delivered packets at the cluster head (CH). The proposed model aims to keep packets transmitting, isotropic or non-isotropic transmission, from the CMs to the CH for maximizing the DT and aims to distinguish between the malicious CM and the one suffering from HW failure. Accordingly, it can manage the consequently lost power due to the malicious attack effect or HW malfunction. Simulation results indicate the proposed mechanism improved performance with TDMA over six different environments against the SF attack that achieves the Pareto optimal DT as compared to a non-cooperative defense mechanism.        


\end{abstract}

\textbf{\textit{Key}words}
Wireless Sensor Networks, Internet-of-things, Game Theory, Threats Mitigation, Power Conservation.


\maketitle

\section{Introduction}
\label{sec:introduction}

According to the dramatic growth of the internet of things (IoT) technologies, IoT systems are utilized in a wide range of applications. In particular, wireless sensor networks (WSNs), which can play a significant role in serving IoT based applications such as smart cities, vehicular networks, environmental and earth monitoring, electrical power lines management, renewable energy adaptation, etc \cite{kim2017smart,lazarescu2013design,fernandes2017internet}. However, WSNs suffer from some weaknesses such as limited power, low processing capabilities, and especially the security and data trustworthiness (DT) aspects. Therefore, WSNs-based IoT security and DT are enormous problems that desire an intelligent and adaptive solution to optimally confront the day-to-day intellectual threats such as selective forwarding (SF), injection, and jamming attacks \cite{hassan2019current,rani2019trust}. 

In the literature context, various trust mechanisms have been proposed to resolve the WSNs security issues. In \cite{butun2013survey}, the authors discussed different intrusion detection systems (IDS), e.g., game theory-based IDS, watchdog-based IDS, cluster-based IDS, etc., to mitigate the WSNs security problems. More particularly, in \cite{shen2018multistage}, game theory has been deployed to establish a valid IDS to realize malware detection infrastructure. In fact, game theory is a dedicated optimization branch that is utilized to handle the interactions of a set of intelligent rational players. More particularly, it aims to enhance their individual payoffs by an intellectual and adaptive manner \cite{ref_5}. More concretely, game theory has emerged due to the distinguishing features in managing the rational players interactions and mitigating several security threats in WSNs such as packet dropping, false data injection, and data delivery corruption \cite{abdalzaher2016game,abdalzaher2017usingiet,abdalzaher2016using,abdalzaher2017effective,abdalzaher2019non,abdalzaher2017using}. However, the above proposals do not consider the hardware (HW) failure which can make the system critically unstable.

The security issue and HW failure (fault) are among the eminent causes of packet dropping in WSNs, which have a severe impact on the DT, network stability, and power consumption. From the security front, dropping packets is a malicious incentive for saving transmission power in clustered WSNs-based IoT. Accordingly, the decision taken can be harmful, which causes a steep degradation for the IoT networks' DT. Moreover, HW failure has several defects. It can cause packet dropping, packet repeating, or over packet transmission. The fault types can be categorized as offset fault, gain fault, stuck-at fault, out of bounds, spike fault, noise fault, data loss fault, redundant data transmission \cite{noshad2019fault}. In {\cite{duche2013sensor}}, the transmission round trip delay of the packets to detect the nodes that suffer from HW failure has been utilized. In {\cite{noshad2019fault}}, the authors extensively studied the HW failure classifier methods such as support vector machine (SVM), machine learning (ML), and random forest. However, most of the presented works in the literature context have only considered the HW faults regardless of the security issue effect and how to distinguish between them. In \cite{abdalzaher2019access}, a Stackelberg game has used to mitigate the corrupted delivered reports in cognitive radio networks due to the selective forwarding data falsification attack in presence of an HW malfunction. However, to the best of our knowledge, no further work has been presented in the literature focusing on classifying between attack effect and HW failure for packet drop. Consequently, an intellectual and adaptive solution is desired to resolve these crucial issues along with the WSNs limited power source.  

In this paper, we propose a game-theoretic approach using a repeated game to enhance WSNs-based IoT security and DT against the SF attack and pinpoint the nodes that suffer from HW failure. The main contributions of the paper are three folds:

\begin{itemize}

\item

The proposed repeated game aims to detect the malicious CMs in clustered WSNs-based IoT due to the SF attack impact, and simultaneously guarantee  that no wrong action is taken even for a node caused by a HW failure at a low probability. The TDMA protocol is used to preserve the synchronization between the CH and CMs, which reduces the detection mechanism complexity and avoids the collision between the delivered packets at the CH. In addition, the isotropic and non-isotropic packet transmissions have been taken in the model consideration.

\item

The proposed approach models a real situation in which some packets can be dropped or over transmitted due to HW failure of the CMs at the presence of SF attack. We verified that the model could improve the DT performance with the presence of HW failure. It can determine the CMs that suffer from HW malfunction among the CMs infected by the SF attack.  

\item

The model attains the Pareto optimality at the optimal DT by resolving the issue of dropping packets due to the SF attack effect. In other words, the model can prevent the designated malicious CMs from dropping packets by supporting these CMs the incentive to act benevolently at which their battery life is conserved. Furthermore, the model can preserve the lost power of packets over transmission or re-transmission as a result of HW failure. The model efficiency is verified through simulation evaluation over six different environments, outdoor line-of-sight (OL), outdoor non-line-of-sight (ON), underground line-of-sight (UL), underground non-line-of-sight (UN), indoor line-of-sight (IL), and indoor non-line-of-sight (IN). Moreover, to draw a realistic WSNs-based IoT system, Tmote Sky mote over the six different environments has been used.

\end{itemize}

The rest of the paper is portrayed as follows. Section \ref{related} presents the related work. In Section \ref{system model}, the system model is discussed. The proposed game approach is then presented in Section \ref{game approach}, while the Pareto optimality and game formulations are addressed in Section \ref{proof}. Section \ref{results} shows the obtained results. Finally, the paper conclusion is revealed in Section \ref{conclusion}.

\section{Related Work}\label{related}

This section discusses the related works of the different WSNs HW failure detection methods and the security front to promote WSNs performance. The security and HW failure in IoT is a prominent problem that can yield into losing data privacy, DT, and wasting power as well. On the one hand, most of the works in the literature context concerning the HW failure utilize the conventional paradigms, e.g., SVM, ML, and random forest by \cite{noshad2019fault,duche2013sensor} and self-diagnosis and cooperative-diagnosis by \cite{jiang2011sensor,jia2018fault,muhammed2017analysis,liu2009fault}. 

In \cite{jiang2011sensor}, the faulty node can be detected using redundant node measurements. Thus, the faulty one is isolated based on its reputation obtained by as a minimum of three neighbor voting. Relying on the neighboring nodes was used to adapt time correlation information between these nodes to detect faulty node/(s) \cite{jia2018fault}. An ineffective solution that can cause more delay and cost that rely on using an extra HW (testbed) to check the faulty node \cite{khalastchi2013sensor}. In \cite{muhammed2017analysis}, three statistical algorithms were studied for fault detection called time-series analysis, descriptive statistics, and Bayesian statistics. The time-series mechanism is utilized to detect packets similarities and to measure the amount of data deviation. Descriptive statistics use the mean or median of neighboring nodes to vote for determining the faulty node. Bayesian techniques were employed to determine the likelihood of a faulty sensor based on Bayes theorem. In \cite{liu2009fault}, the authors mentioned that the node could detect the failure by self-diagnosis, such as the faults caused by battery depletion, which is measured by the battery current or voltage. Accordingly, if a node dropped or lost packets, an extra memory should be embedded in the nodes to retrieve those packets and resend again-they said. The authors also respite the lost packet to the environment condition that is why we study the proposed paradigm over six different environments.

On the other hand, uncontrolled WSNs security can deteriorate the DT, data privacy, and power consumption. In this trajectory, game theory has introduced several approaches \cite{abdalzaher2016game,salim2016energy,abdalzaher2016using,abdalzaher2019non,abdalzaher2017effective,abdalzaher2017usingiet,abdalzaher2017using,sfar2019game}. In \cite{abdalzaher2016game}, the authors extensively studied various WSN attacks and the suitable game defense models. In \cite{sfar2019game} Markovian chain was utilized to model the game transitions for preserving the data privacy. In \cite{abdalzaher2016using,abdalzaher2017usingiet}, Stackelberg games have been developed to mitigate the external attacks manipulations using energy defense to avoid the delivered data disruption in a clustered WSN. Stackelberg game has also been extended to confront the false injected data from intelligent attacks in WSNs to enhance the DT \cite{abdalzaher2017effective}. The Stackelberg game was also utilized to confront the false injected noise power in WSNs-based cognitive radio (CR) due to the spectrum sensing data falsification (SSDF) attack, where the HW failure problem was considered \cite{abdalzaher2019access}. The work in \cite{abdalzaher2019access} aims to mitigate the disrupted observed signal-to-noise ratio due to the SSDF attack to make the fusion center in CR capable of achieving accurate decision about the spectrum status. Interestingly, in \cite{abdalzaher2019non}, a nonzero-sum game was proposed to mitigate DoS attack and ON-OFF attack impact and to detect the HW failure in WSNs. The SF attack was handled by game theory in \cite{abdalzaher2016using}. In \cite{abdalzaher2017using}, a repeated game model has been utilized to enhance WSNs DT against the SF attack. More particularly, the most effective parameter in that game was the distance between the communicating nodes, which was criticized by \cite{zhou2004impact}. Moreover, the model in \cite{abdalzaher2017using} omitted the harmful effect of HW failure and did not utilize the TDMA protocol as well. To solve this issue, a more intelligent game model needs to be designed along with a robust detection scheme against the SF attack at the HW failure existence.

Unlike the previous related works, this paper presents an efficient repeated game-theoretic approach along with TDMA protocol to classify between the cause of packet dropping in WSNs-based IoT whether a reason of malicious effect due to SF attack or a result of HW failure to manage the consequent power waste and achieve the optimal DT. The repeated game is considered the best among the cooperative games to support the adequate incentive to guarantee collaborative interaction between the participant players at the price of elapsed time till reaching the equilibrium point, which meets the flexibility of IoT applications time constraint. The proposed model presents the adequate incentive to the malicious CMs due to SF attack to react benevolently and then, uncomplicatedly at the equilibrium point, detects the CMs suffering from HW failure. \vspace{-.1in}

\section{System model}\label{system model}

\begin{figure}[h]
\begin{center}
\includegraphics[width =\columnwidth]{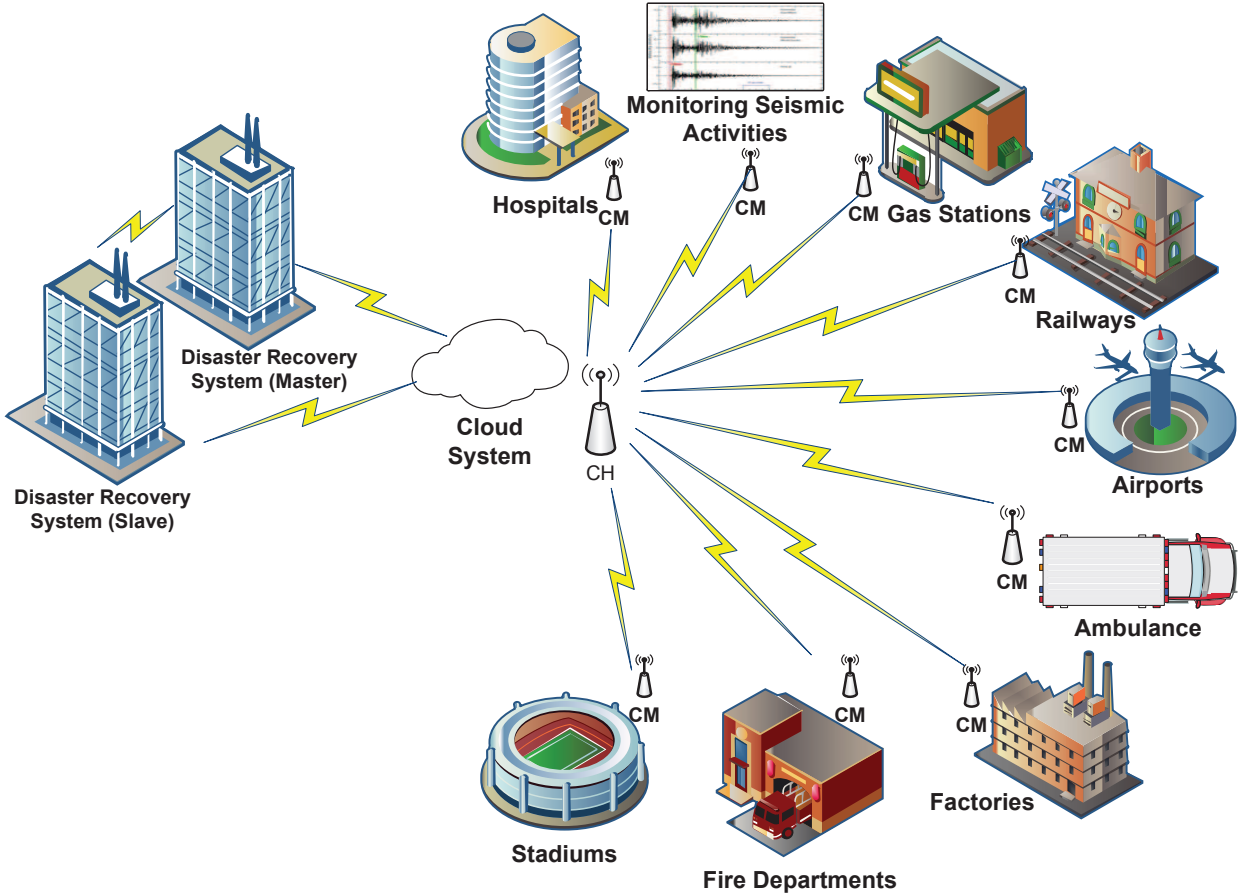}
\caption{Disaster management using WSNs-based IoT system.}
\label{disaster_1}
\end{center}
\end{figure}

\begin{figure}[h!]
\begin{center}
\includegraphics[width =\columnwidth ]{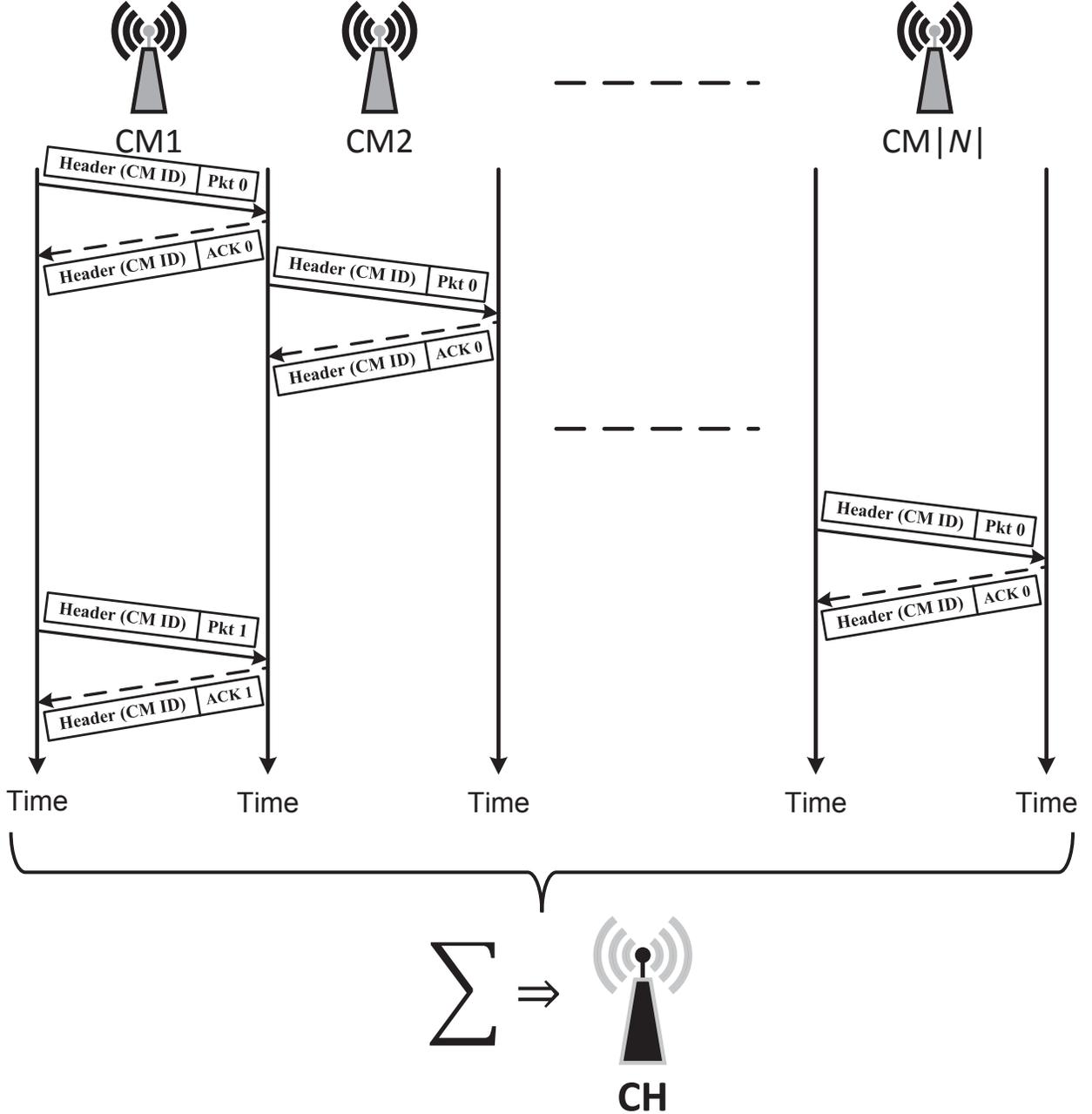}
\caption{Packet transmission sequence using TDMA protocol.}
\label{Time}
\end{center}
\end{figure}

Figure \ref{disaster_1} illustrates system model using IoT based WSN. In this model, WSN consists of a set of CMs, $\mathcal{N}$, where the number of CMs in the cluster is given by the cardinality of the set $\mathcal{N}$ which is represented by $\mathcal{|N|}$. Table \ref{list} lists the utilized notations and variables. For managing synchronization and packet transmission between the CMs and CH in the clustered WSN, TDMA protocol is utilized as shown in Fig. {\ref{Time}}. In fact, the proposed model is applicable to work with different MAC protocols such as OFDM based protocols \cite{abdalzaher2017using} but we utilize TDMA due to its features supporting the power conservation. As extensively studied in the literature, TDMA is effective for prolonging the network lifetime \cite{van2004advantages}. Moreover, dynamic TDMA, which is a MAC protocol based on scheduled time (DMAC), is efficient for WSNs energy conservation, prolonging network lifetime, and avoiding overhearing \cite{lu2004adaptive}. Consequently, it is an adequate intensive to employ TDMA protocol with the proposed approach.  

This paper considers a game-theoretic model where each game player has two actions status. First, every $i$-th CM has to execute either benevolent action by not dropping packets, no drop ($ND$); or malicious action by dropping packets, drop ($D$). The $D$ action is an incentive of the CM to save battery power. Second, the CH performs Beacon ($B$) or no Beacon ($NB$) action status. The $B$ action means that permission is given to the benevolent CM, does not drop packets, to send their observed data. In addition, $B$ is used to permit the $i$-th benevolent CM to go to the sleep mode, to take a rest of packet transmission to save its battery lifetime, or to get power recovery when recharging power is available. Conversely, the $NB$ action is fundamentally used when the $i$-th CM action is $D$. Moreover, a controversial CM performance occurs when it has a random HW failure at the same time of SF attack exist; specifically, the ones suffering from dropping packets and over packets transmissions which are precisely considered through the proposed model. In other words, the model can effectively classify between the infected CMs by the SF attack and the ones suffering from HW failure. The TDMA is also utilized to gradually send the $B$ to the benevolent CM when its turn comes to transmit its data.


\begin{table}[h!]
\centering
\caption{Notations of parameters and variables.}
\label{list}
\begin{tabular}{|c | l |}
\hline 
 Symbol & Description \\
  \hline \hline
   $\mathcal{N}$ and $|\mathcal{N}|$ & The set of CMs and the total number of CMs\\
    \hline 
    $A_{i}$ & An individual action of $i$-th CM \\
    \hline
    $A_{i}^*$ & The optimal action of $i$-th CM \\
    \hline
    $\mathcal{\textbf{A}}_i$ & Set of action done by the $i$-th CM \\
    \hline    
        $A_{CH}$ & An individual action of the CH \\
    \hline
    $A_{CH}^*$ & The optimal action of the CH \\
    \hline
    $\mathcal{\textbf{A}}_{CH}$ & Set of action done by the CH\\
    \hline
    $\textbf{A}$ & The action matrix  \\
    \hline
    $B$ and $NB$ & \begin{tabular}[c]{@{}l@{}}Beacon and No Beacon actions done by \\the CH, respectively\end{tabular}  \\
    \hline
    $D$ and $ND$ & Drop and No drop actions done by the CM, respectively  \\
    \hline
    $RSSI$ & Received signal strength indicator  \\
    \hline
    $TC$ & Transmission cost \\
    \hline
    $TC_0$ & Start up initial cost \\
    \hline
    $TC_A$ & Amplification cost  \\
    \hline
    $V$ & Battery voltage  \\
    \hline
     $T_0$ & Offset (initial) start up time  \\
    \hline

     $I_0$ & Consumed current when radio is ON\\
    \hline
  
    $I_c$ & Current supported by the battery at transmission level $c$  \\
    \hline
    $DR$ & The transmission data rate  \\
    \hline
    $L$ & The packet length in bits\\
    \hline
    $PL$ & \begin{tabular}[c]{@{}l@{}} The path loss of the link between every $i$-th CM\\ and CH\end{tabular} \\
    \hline
    $PL_{f}$ & \begin{tabular}[c]{@{}l@{}} The free space path loss at the reference distance of the \\antenna far field \end{tabular}\\
    \hline
    $d_{iCH}$ & The distance between the communicating CM $i$ and CH  \\
    \hline
    $d_{0}$ & The reference distance of the antenna far field   \\
    \hline
    $h$ & Transmission power level  \\
    \hline
    $n$ & The path loss exponent  \\
    \hline
    $\sigma$ & The standard deviation of the shadow fading  \\
    \hline
    $\eta$ & The path loss direction coefficient\\
    \hline
$\theta$ & The radiation coefficient angle  
      \\
      \hline
      $R$ & A random value $\in ]0,1[$\\
    \hline    
     $DOI$ & Degree of irregularity\\
    \hline
    $RL$ & The reliability level\\       
    \hline   
    $TP$ & \begin{tabular}[c]{@{}l@{}}The total number of packets per message\\ (a given window size) \end{tabular}\\    
    \hline    
    $pkt$ & Data packet   \\
    \hline
    $U_i$ & The The utility function of the $i$-th CM\\ 
    \hline    
    $\xi_i$ & \begin{tabular}[c]{@{}l@{}}A punishment (loss) parameter deducted \\from the $i$-th CM utility \end{tabular}\\
    \hline    
    $\alpha \; \text{and} \; \beta$ & Application dependent weighting factors\\
    \hline
    $eb$ & Transmission energy per bit\\
    \hline
     $DT$ & The data trustworthiness\\
    \hline
    $\overline{DT}$ & The optimal data trustworthiness\\
    \hline
     $rd$ & The round number\\
    \hline
     $N_{rd}$ & The total number of rounds\\
    \hline
\end{tabular}
\end{table}

\begin{figure}[t!]
\begin{center}
\includegraphics[width =\columnwidth]{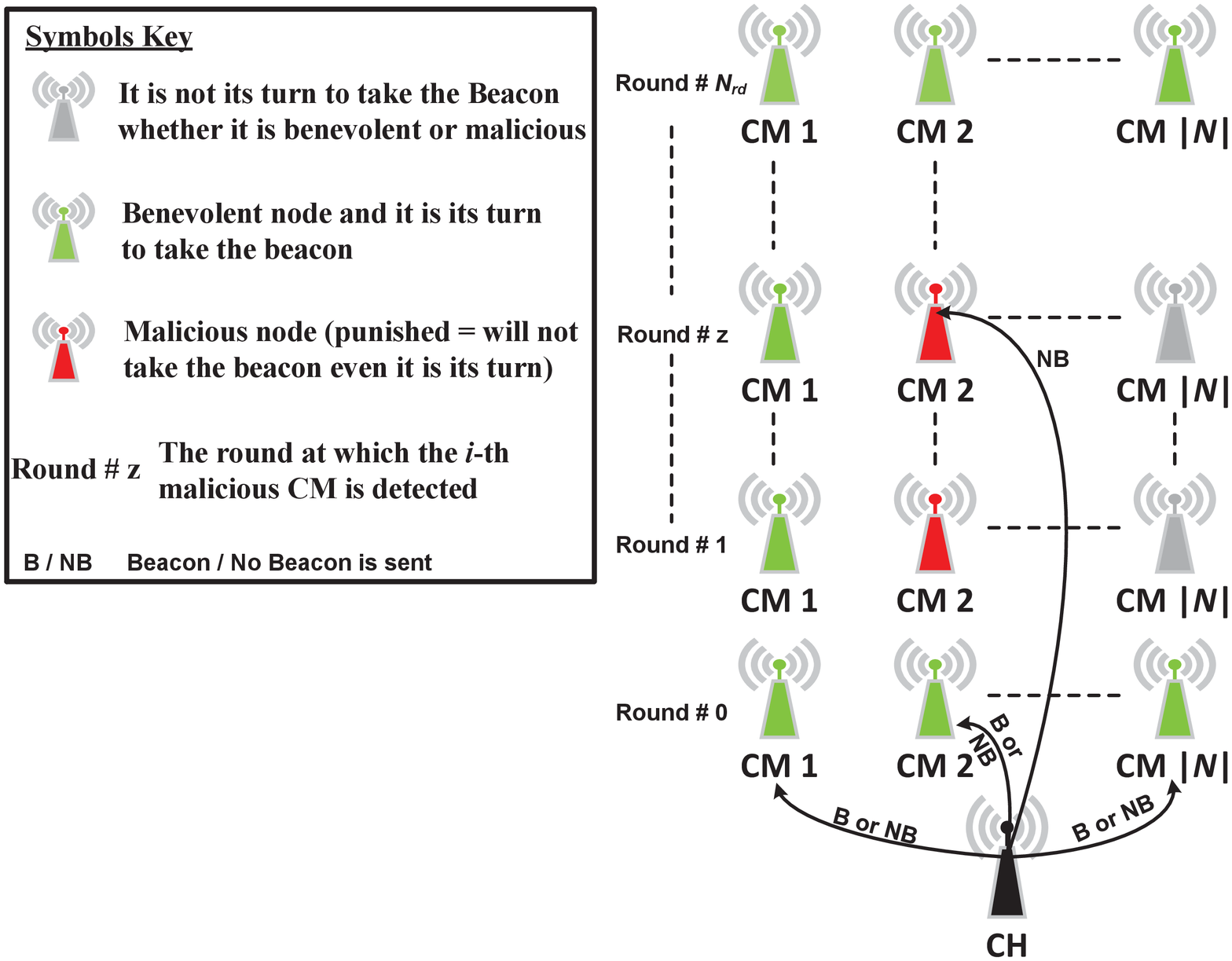}
\caption{CH Beacon distribution for the CMs with TDMA.}
\label{beacon}
\end{center}
\end{figure}

Figure {\ref{beacon}} shows the Beacon ($B$) action/message distribution after checking the CM behavior. At the beginning (Round $\#$ 0), all CMs are supposed to be benevolent. Accordingly, the CH action for all CMs is $B$. After calculating ($U_i$)s for all CMs (at Round $\#$ z), the CH can determine which CM has the right to be supported by the Beacon or not. Therefore, if the $i$-th CM is benevolent and it is its turn using TDMA, it will receive the Beacon. On the contrary, if this $i$-th CM is malicious (drops packets), the Beacon will not be sent to this $i$-th CM. Accordingly, this malicious CM will keep re-transmission of the packets that will not receive acknowledgments on from the CH. Therefore, this approach can deteriorate the battery power of the CM that keeps on malicious behavior, and hence, this CM is prone to going to die. To this end at (Round $\# \; N_{rd}$), $N_{rd}$ is the given period representing the total number of rounds, all CMs are benevolent, and no one will be changed after that unless some of them have HW failure. Indeed, the CH communicates with the CMs throughout a star topology, in which the communication link between each CM and the CH is dedicated, to facilitate the check procedure of the received packets from every $i$-th CM. Therefore, it will be a conventional topology to determine the benevolent CM that does not drop. To this end, there are four scenarios in order due to the CH action ($A^i_{CH}$) and the $i$-th CM action ($A_{i}$). Three scenarios out of these four represent the one-shot games at which both the CH and CM perform the same action regardless of the other player's action. The last scenario introduces the rational interaction between the CH and the $i$-th CM as a reward and punishment defense mechanism according to the action done by every CM. Based on this scenario, all the malicious rational CMs will re-behave benevolently. Finally, the four actions status of the two players (i.e., CH and the $i$-th CM) are given by

\begin{itemize}
\item
$A^i_{CH} =$ $NB$ and $A_i =$ $D$
\item
$A^i_{CH} =$ $NB$ and $A_i =$ $ND$
\item
$A^i_{CH} =$ $B$ and $A_i =$ $D$
\item
$A^i_{CH} =$ $B$ and $A_i =$ $ND$ 
\end{itemize}  


\section{Proposed Repeated Game Approach}\label{game approach}

The proposed game confirms a cooperative attitude between the CH (the defender) and the participant CMs. Indeed, the CMs are assumed to act as rational players. The model is used to mitigate the potential malicious CMs that drop the observed packets due to the effect of an intelligent SF attack. The proposed model provides an adequate incentive for the CMs to react benevolently and not drop packets. Conversely, if a CM persists to behave maliciously, this CM will not be given a permission to go to the sleep mode leading to shortly exhausting its battery.


Consequently, rational CMs will prefer to react benevolently because it is more profitable based on the proposed game. This game leads to the Pareto optimality and nash equilibrium (NE) at which both players reach their optimal utility. The utility function of every $i$-th CM per every iteration is computed using three main parameters, namely, received signal strength indicator ($RSSI_{i}$), reliability level ($RL_{i}$), and a punishment parameter ($\xi_i$) between the CH and the $i$-th CM, which is given by 
\begin{equation}\label{utility}
U_{i} = \alpha RSSI_{i} + \beta  RL_{i} - \xi_i,
\end{equation}\vspace{-.2in}
\begin{equation}\label{eq2.9}
 \alpha + \beta   = 1,
\end{equation}
where $\alpha$ and $ \beta$ are ''application dependent'' weighting factors \cite{abdalzaher2017using}.

\subsection{Transmission Cost based Received Signal Strength Indicator ($RSSI$)}
First, this parameter concentrates on the main effective factors of the transmission cost that are described as follows. 
\begin{equation}\label{rssi}
RSSI = TC - PL - P_n,
\end{equation}
where $TC$ is the transmission cost while $PL$ is the path loss and $P_n$ is the noise power that are obtained from the six exploited environments representing the link quality indicator (LQI) that the CM is deployed over. $P_n$ and $PL$ parameters measurements are indicated in Table {\ref{pathloss}} \cite{gungor2010opportunities}. Then, the transmission cost is expressed by   
\begin{equation}
TC = TC_0 + TC_A,
\end{equation}
where $TC_0$ is the initial cost that the CM starts up with, while $TC_A$ denotes the amplification cost for packet transmission. The initial cost can be given by
\begin{equation}
TC_0 = V \times I_0 \times T_0,
\end{equation}
where $V$ is the battery voltage exploited for transmission, $I_0$ represents the current in Amperes when the radio is in (ON state), and $T_0$ is the initial start up time.
\begin{equation}
TC_A = V \times I_c \times \frac{L}{DR},
\end{equation}
where $I_c$ is the current in Amperes with a transmission power level $c$, $L$ represents the packet length, and $DR$ is the transmission data rate. Table \ref{tmote} shows the eight transmission levels along with the corresponding consumed electric current using Tmote Sky mote \cite{corporaton2006tmote}. The adopted path loss between every $i$-th CM and the CH ($i,CH$) is given by

\begin{table}[]
\centering
\caption{Path loss parameters \cite{gungor2010opportunities}.}
\label{pathloss}
\begin{tabular}{|l|l|l|l|}
\hline
Environment                        & $n$  & $\sigma$ (dB) & $P_n$ (dBm) \\ \hline
Outdoor line-of-sight (OL)         & 2.42 & 3.12 &   -93      \\ \hline
Outdoor non-line-of-sight (ON)     & 3.51 & 2.95  &    -93         \\ \hline
Underground line-of-sight (UL)     & 1.45 & 2.45   &   -92         \\ \hline
Underground non-line-of-sight (UN) & 3.15 & 3.19   &   -92         \\ \hline
Indoor line-of-sight (IL)          & 1.64 & 3.29   &   -88         \\ \hline
Indoor Non-line-of-sight (IN)      & 2.38 & 2.25  &    -88         \\ \hline
\end{tabular}
\end{table}

\begin{figure}[t!]
\begin{center}
\includegraphics[width =\columnwidth]{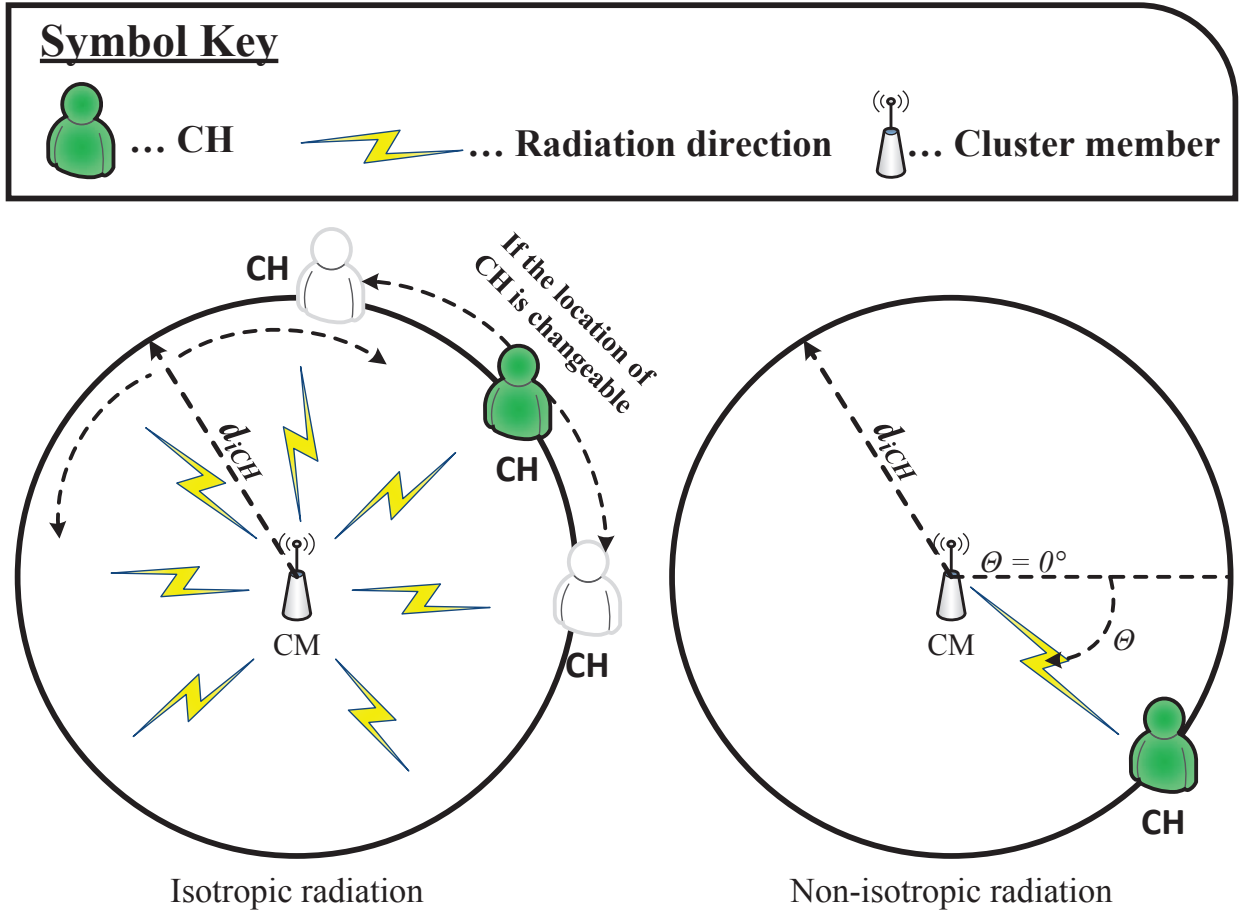}
\caption{Isotropic and Non-isotropic radiation.}
\label{isotropic}
\end{center}
\end{figure}

\begin{table}[]
\centering
\caption{Tmote Sky transmission levels and consumed electric current.}
\label{tmote}
\begin{tabular}{|l|l|}
\hline
$h$ & Electric current consumed (mA)  \\ \hline
3   & 8.5                            \\ \hline
7   & 9.9                           \\ \hline
11  & 11.2                           \\ \hline
15  & 12.5                               \\ \hline
19  & 13.9                            \\ \hline
23  & 15.2                          \\ \hline
27  & 16.5                          \\ \hline
31  & 17.4                            \\ \hline
\end{tabular}
\end{table}  



\begin{equation}
PL[\texttt{dB}] = \bigg(PL_f[\texttt{dB}] + 10n \log10 \frac{d_{iCH}
}{d_0} \bigg) \times \eta_{\theta} + \sigma[\texttt{dB}],
\end{equation}
where $PL_{f}$ is the free space path loss at the reference distance $d_0$ of the antenna far-field, $n$ denotes the path loss exponent, $d_{iCH}$ is the distance between the transmitting $i$-th CM and the CH, $\sigma$ is the standard deviation in dB of the shadow fading (log-normal distribution) measurements as indicated in Table {\ref{pathloss}}, $\eta_{\theta} \in ]0,1]$ is the path loss direction coefficient when non-isotropic radiation is used, and ${\theta}$ represents the  radiation coefficient angle which is depicted in Fig. \ref{isotropic}.

If the radiation is isotropic, $\eta_{\theta} = 1$. On the other hand, when the radiation is non-isotropic $\eta_{\theta}$ is expressed by
\begin{equation}\label{eq2.4}
\hskip-10pt \eta_{\theta}=
\begin{cases}
 1, & \text{if}\ \ \theta = 0 \\
R \times DOI, & \text{if}\ \ 0^{\circ} < \theta < 360^{\circ},
\end{cases}
\end{equation}
where $R$ is a random value $\in ]0,1[$ and the experienced values of degree of irregularity (DOI) are obtained from the experiments presented in \cite{zhou2004impact}.

\subsection{Reliability Level ($RL$)}

Reliability level for the $i$-th CM ($RL_{i}$) represents the percentage of transmitted packets from the total number of packets ($TP$) per message (i.e., window size) during a given period, which is given by the CH as
\begin{equation}\label{RL}
RL_{i} = \frac{\sum_{j=1}^{TP} f(pkt^{j}_{i})}{TP},
\end{equation}
\begin{equation}\label{eq2.4}
f(pkt^{j}_{i})=
\begin{cases}
1, & \text{if}\ \ \textit{j}\text{-th} \   \text{pkt is forwarded from \textit{i}\text{-th}} \\ & \text{CM to CH}\\
0, & \text{if}\ \ \textit{j}\text{-th} \   \text{pkt at \textit{i}-th CM is dropped} \\ & \text{(not forwarded to CH)},
\end{cases}
\end{equation}
where $pkt_i^{j}$ corresponds to the $j$-th forwarded packet (pkt) from $i$-th CM to the CH, and $RL_{i}$ $\in ]0,1]$. 

\subsection{Punishment Parameter ($\xi$)}

The punishment parameter for the $i$-th CM $\xi_i$ in Eq. {\ref{utility}} is utilized to enhance the cooperation between the CH and the participant CMs, which is calculated by
\begin{equation}\label{xi}
\hskip-10pt \xi_i=
\begin{cases}
x1, & \text{if}\ \  A^i_{CH} = NB \ and \ A_i = ND\\
x2, & \text{if}\ \ A^i_{CH} = B \ and \ A_{i} = D\\
x3, & \text{if}\ \ A^i_{CH} = NB \ and \ A_i = D\\
0, & \text{if}\ \ A^i_{CH} = B \ and \ A_i = ND,
\end{cases}
\end{equation}
where $x1$, $x2$, and $x3$ are given by
\begin{equation}\label{x1}
x1 = \sum^{TP}_{j=1} f(pkt^{j}_i) \times \text{L} \times eb,
\end{equation}
\begin{equation}\label{x2}
x2 = \bigg(TP - \sum^{TP}_{j=1} f(pkt^{j}_i)\bigg) \times \text{L} \times eb,
\end{equation}
\begin{equation}\label{x3}
x3 = x1 + x2,
\end{equation}
where $eb$ denotes the transmission energy per bit, and $TP$ is the total number of packets to be transmitted to the CH.

In fact, the term $\xi_{i}$ in Eq. (\ref{utility}) is a loss parameter used to stimulate both CMs and CH to establish a cooperative connection leading to prolonging the network lifetime where three main $\xi_{i}$ cases are used for. First, the CH does not send Beacon message to the $i$-th CM; however, the CM is benevolent by keeping packets transmission to the CH. Therefore, this CM cannot take a rest of packet transmission or go to the sleep mode for saving the battery lifetime at which $\xi_{i} = x1$ due to the non-cooperative behavior of CH. Second, $\xi_i$ equals $x2$ when the CM behaves maliciously by dropping packets; however, the CH cooperates with it by sending it the Beacon. Third, the worst-case at which both the CH and CM choose to be non-cooperative players for which $\xi_i$ equals $x3 = x1 + x2$.       

\subsection{Data Trustworthiness (DT)}
According to the above-mentioned subsections, the DT calculated by the CH is defined as the average utility function \cite{abdalzaher2017using} of all CMs during the given time period as 
\begin{equation}\label{DT}
DT = \frac{\sum^{N_{rd}}_{rd=1} \sum_{i=1}^{\mathcal{|N|}}U_{i}^{rd}}{N_{rd}}.
\end{equation} 
It is worth mentioning that the relationship of the packet drop rate and the DT is inversely proportional.

\subsection{The Proposed Model and Corresponding One Shot Games}

The proposed repeated game is utilized to resolve the equivalent \textit{prisoner-dilemma} \cite{ref_5} using the available four scenarios based on the players' actions. The game is developed to fulfill the Pareto optimal point and NE point at the same state. The first three scenarios represent a one-shot game of \textit{prisoner-dilemma} between the CH and the $i$-th CM in which the DT does not reach to its optimal value. The last scenario presents the proposed repeated game versus the other one-shot games, which are discussed as follows:

\begin{itemize}

\item  In the first scenario, all CMs actions are to cooperate with the CH by not dropping any packet ($ND$). In other words, they behave benevolently with the CH. On the other hand, the CH action is $NB$ to be sent to the CMs even they are benevolent. Accordingly, the utility function is affected by Eq. (\ref{x1}) of the punishment parameter.

\item  In the second scenario, the CH chooses to exert $B$ action for all CMs. Conversely, the CM action is to drop ($D$) packets aiming at saving battery lifetime. Therefore, the punishment parameter here is represented by Eq. (\ref{x2}).   

\item The third scenario is the worst one at which the minimal utility is obtained due to the CH and CM actions, $NB$ and $D$, respectively. It means that the CH does not provide the CM by the Beacon, and this CM action is to drop packets. Consequently, the punishment parameter ($\xi$) is calculated based on Eq. (\ref{x3}). Therefore, these malicious CMs will not be permitted to go to sleep mode, and hence, they are not capable of preserving their battery lifetime.

\item  The fourth scenario represents the proposed repeated game model for handling the behavior of the $i$-th malicious CM. At the equilibrium point, the CH and all $i$-th CM choose to cooperate by exerting $B$ and $ND$ actions, respectively. More particularly, both RL and utility functions are maximized. Generally speaking, the process of permitting or preventing a CM from the Beacon message until reaching the optimal DT is based on (\textbf{Algorithm 1}). In this algorithm, if a CM drops packets, it will be designated as malicious or suffers from HW failure. In other words, the dropping packets CM will be designated as malicious or labeled in the HW failure list (HWL). Then, this $i$-th CM will go through a punishment cycle till reacting benevolently. At satisfying two constraints, the round number $ rd = i + i\mathcal{|N|}$ comes and this $i$-th CM reacts benevolently again, this $i$-th CM will be given the Beacon. For instance, if $\mathcal{|N|}$ equals 10 and CM 3 is malicious, the Beacon will be supplied to this CM only when it re-behaves benevolently at $rd = 33$. This scheme is recursively executed $N_{rd} = c \mathcal{|N|}$ times till achieving the optimal DT, where $c \in \mathbb{R}_{+}>\mathcal{|N|}$. Consequently, when the CM action is $ND$, its gain is much higher than when its action is $D$ for its utility function. Similarly, when the CH action is $B$ to the CMs, their utility functions are enhanced, which yields to the optimal DT. Therefore, this process is called the punish and forgive strategy. Finally, both the CH and the designated rational malicious CM based on this game will prefer to cooperate due to the gain they can earn. In other words, the cooperative interaction between CH and CMs leads the Pareto optimality and NE at the same state by reaching the optimum DT. After that, any CM drops packets or does over packet transmission is a result of HW failure, which is listed in HWL and quietly detected by the CH.  
\end{itemize}

\begin{algorithm}[t]
\caption{Proposed Repeated Game Algorithm for SF attack. 
}
\SetAlgoLined
 \textbf{Input} $A^i_{CH} = \{B, NB\}$, $A_i = \{D,ND\} $ \;

 \While{$ d \leq c\mathcal{|N|}$}{
  Compute $RSSI_i$, $RL_i$, $\xi_i$, and $U_{i}$ by Eqs. (\ref{rssi}), (\ref{RL}), (\ref{xi}) and (\ref{utility}) $\forall i \in |\mathcal{N}|$\;

  Determine the malicious CMs or the ones have HW failure\;
 
	{\eIf{ CM$_i$ drops pkt $\rightarrow$ A$_i$ is $D$}
	{This $i$-th CM $\rightarrow$ malicious/HWL\;}
	{This $i$-th CM is designated as benevolent and No HW failure;}}
	
   {\eIf {$A_i$ of this $i$-th CM is  $ND$ and $rd = i + i\mathcal{|N|}$}
   {This $i$-th CM will receive $B$ message ($A^i_{CH}$ $\rightarrow B$)\;}
   {This $i$-th CM will be listed in HWL;}}
{\eIf{DT is calculated using Eq. (\ref{DT}) and ($A^i_{CH},A_i$) $\rightarrow$ ($A_{CH}^*,A_i^*$) $\forall i$}
{DT is optimal and $NE$ exists\;}{Continue\;}}

}
 \textbf{Output} Optimal DT is attained and $(A^*_{CH}, A^*_i), \forall i$ $\rightarrow$ NE exists\; 
\If{$A_i = D$}
{This $i$-th CM suffers from a HW failure\;} 
\end{algorithm}


\section{Pareto Optimality and Game Formulations}\label{proof}

The proposed repeated game is between the CH (defender) and every $i$-th CM, which is denoted by
\begin{equation}
G=\langle \{CH, i\}, \textbf{A}, U_{i}\rangle.
\end{equation}

The CH set of actions $\mathcal{\textbf{A}}_{CH}$ represent the CH defense strategy $DS_{CH}(\xi_i, DT)$ against the designated malicious CM. Conversely, $\mathcal{\textbf{A}}_{i}$ denotes the $i$-th CM malicious strategy $MS_{CM}(f(pkt^{j}_i), DT)$ at which the $j$-th packet is dropped by the $i$-th CM. 

The best strategy is achieved using the proposed game represented by the fourth scenario in which the optimal action of the $i$-th CM $A^*_i \in \mathcal{\textbf{A}}_i$ is given by
\begin{equation}\label{eq2.11}
A^*_i = \underset{\xi_i=0}{\operatorname{arg\,max}} (U_i).
\end{equation}

More concretely, the optimal action of the $i$-th CM and CH are given by 
\begin{equation}\label{eq2.12}
A^*_i = ND \rightarrow f(pkt^{j}_i)= 1, \forall j \in TP,
\end{equation}

\begin{equation}\label{eq2.13}
A^*_{CH} = \underset{f(pkt^{j}_i)= 1, \forall j \in TP, \forall i \in \mathcal{N}, j \neq i}{\operatorname{arg\,max}} (DT).
\end{equation}

\noindent Taking into consideration the rationality principle of the participant players, the CH will choose to cooperate and use $B$ action. Consequently, $A^*_{CH}$ can be simplified as 
\begin{equation}\label{eq2.14}
A^*_{CH} = B \rightarrow  \xi_{i} = 0, \forall i.
\end{equation} 

Consequently, the NE is achieved, which is given here by
\begin{equation}\label{eq2.15}
U_i(A^*_{CH},A^*_i) \geq U_i(A^i_{CH},A^*_i), \forall i \in \mathcal{N}.
\end{equation}

Then, we prove that both the Pareto optimal and NE point are achieved in the same state as follows:

Let the Pareto optimality is attained at optimal DT ($\overline{DT}$). Consequently, any other value of DT is less than this optimal value as 

\begin{equation}\label{eq8.16}
\overline{DT} \geq DT.
\end{equation}

$\overline{DT}$ is attained using the punish-and-forgive strategy that is presented in the fourth scenario, which is derived by 
 
\begin{equation}\label{eq8.17}
\overline{DT} = \quad \min_{A_{CH}} \max_{A_{i , \forall i \in \mathcal{N}}} DT,\
\end{equation}

\begin{equation}\label{eq8.18}
\quad \min_{A_{CH}=B} \xi_i = 0\, \; \forall i \in \mathcal{N},\
\end{equation}

\begin{equation}\label{eq8.19}
\therefore U_{i} = \alpha RSSI_i + \beta RL_i, \forall i.
\end{equation}

\begin{equation}\label{eq8.20}
 \quad \max_{A_{i, \forall i \in \mathcal{N}}}  f(pkt_i^{j}), \forall j \in TP, \; \forall i \in \mathcal{N}, j \neq i,\
\end{equation}

\hspace{-15pt} Accordingly, $A^*_i = ND \rightarrow f(pkt_i^{j}) = 1, \forall i \in \mathcal{N},  j \neq i,$ and $ A^*_{CH} = B \rightarrow \xi_i = 0, \forall i \in \mathcal{N}$

\begin{equation}\label{eq8.21}
\therefore RL_i = 1, \forall i \in \mathcal{N},
\end{equation}

\begin{equation}\label{eq8.22}
\therefore \overline{DT} > DT.
\end{equation}

Consequently, the NE and Pareto optimality are achieved at the same state when $\overline{DT}$ exists at $(A^*_{CH},A^*_i), \forall i \in \mathcal{N}$.

\section{Simulation results and Discussion}\label{results}

This section shows the simulation results. The simulation parameters are depicted in Table \ref{parameters}. We use Tmote Sky mote as a realistic node model among the most usable nodes with standard of IEEE 802.15.4 supporting precise locations and scalability for IoT platforms \cite{raza2011securing} that equipped by the Texas Instruments MSP430 microcontroller with a Chipcon CC2420 radio, and utilize the parameters from its datasheet in \cite{corporaton2006tmote}. 
We also assumed that 20\% of CMs suffer from a random HW failure, and all CMs use power level $h=31$ as an empirically selected optimum value. The utilized DOI measurement values are $\{0.0055, 0.0035,0.004,0.0045,0.006,0.0085\}$ \cite{zhou2004impact}. 
All CMs are located at the maximum coverage range between the CH and every Tmote Sky CM (125 m) to show the effectiveness of the proposed model. Besides, the DT of the received signal is affected by $P_n$, path loss, and shadowing fading (log-normal distribution with standard deviation $\sigma$). We assume six environment scenarios (i.e., OL, ON, UL, UN, IL, and IN), where their path loss parameters are given in Table {\ref{pathloss}} \cite{gungor2010opportunities}. Rayleigh fading is not taken into consideration. 


\begin{table}[b]
\caption{Simulation parameters.}
\label{parameters}
\centering
\begin{tabular}{|l|l|l |l|}
\hline
Parameter       & Value     & Parameter   & Value                                                                                                \\ \hline \hline
$\mathcal{|N|}$ & 10        & \begin{tabular}[c]{@{}l@{}} No. of randomly assumed \\malicious CM \end{tabular} & 1 and 2                                                 \\ \hline
$L$        & 1024 bits     & $T_0$       & 580 $\mu$s                                                                                              \\ \hline
$I_0$        & $\mu$A   & $DR$   & 250 Kb/s                                                                                             \\ \hline
$I_c$        & 17.4 mA      & $V$   & 3 Volts \\ \hline
$\alpha$ and $\beta$        & 0.6 and 0.4    & $eb$   & 50 nJ \\ \hline
 $d_0$      & 10 m \cite{corporaton2006tmote} & $d_{iCH}$  & 125 m \cite{corporaton2006tmote}  \\ \hline
$PL_f$             & 55 dB   & $h$             & 31 \\ \hline
$TP  $           & 100  & $c$   & 11 \\ \hline
\end{tabular}
\end{table}

\begin{figure}[t]
\begin{center}
\includegraphics[width =.65\columnwidth]{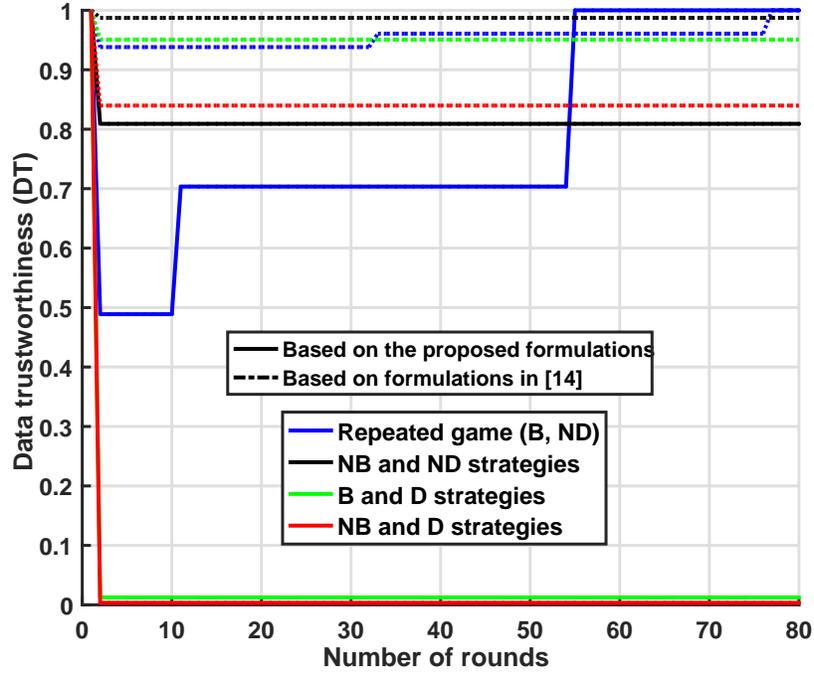}
\caption{Normalized DT comparison between the proposed model and the corresponding in the literature using Non-isotropic transmission over UL environment, number of malicious CMs equals 2.}\vspace{.3in}
\label{comp_old}
\end{center}
\end{figure}

Figure \ref{comp_old} illustrates the improved performance using the proposed game as compared to the work in \cite{abdalzaher2017using}. It is clear that the normalized DT of the proposed game reaches the equilibrium state much faster than the corresponding game model in \cite{abdalzaher2017using}. Moreover, the obtained normalized DT of the three one-shot games involved in this paper gets worse as compared to the proposed repeated game and corresponding results in \cite{abdalzaher2017using}. It means that the proposed repeated game is more effective than the work in \cite{abdalzaher2017using} for stimulating the CH and CM to cooperate using $B$ and $ND$ actions, respectively.

\begin{figure}[h!]
\begin{center}
\includegraphics[width =.65\columnwidth]{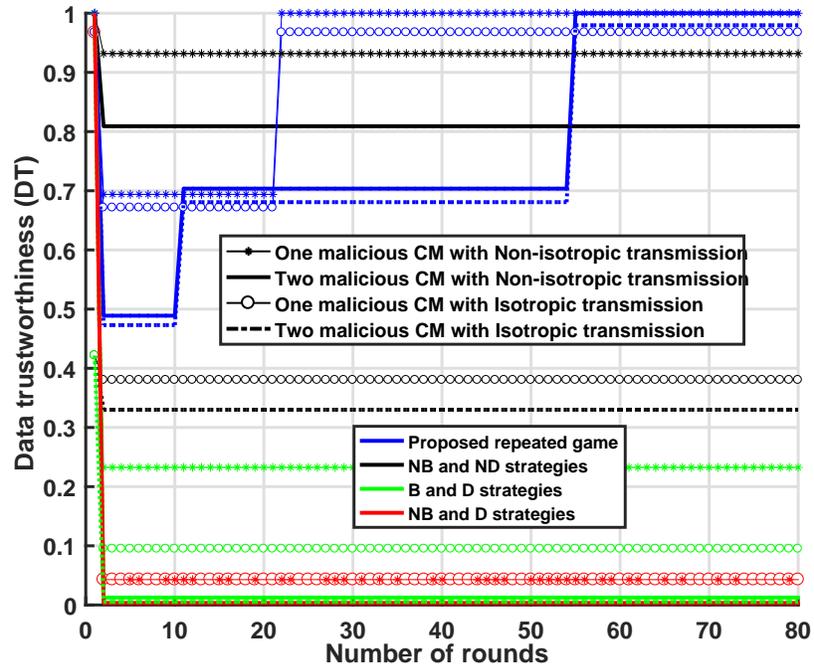}
\caption{Normalized DT based on the proposed repeated game compared to the three one shot games using Isotropic and Non-isotropic transmission and over ON environment, when number of malicious CMs equals 1 and 2.}\vspace{.2in}
\label{model_3shotG}
\end{center}
\end{figure}

Figure \ref{model_3shotG} shows the obtained results of the proposed repeated game model as compared to the three one-shot games. It is clear that after the equilibrium point, the model maximizes the normalized DT leading to the optimal DT whether the number of malicious CMs equals 1 or 2. We succeeded to prove that in both the isotropic and non-isotropic packet transmission. The minimum normalized DT is achieved by the third scenario when the CH and CM actions are $NB$ and $D$, respectively. This is because the punishment parameter is considered doubled using Eq. (\ref{x3}). The normalized DT of the first and second one-shot games (first and second scenarios) get better performance than the third one but are still less than the proposed game results.

\begin{figure*}[!ht]
    \centering 
\begin{subfigure}{0.3\textwidth}
  \includegraphics[width=\linewidth]{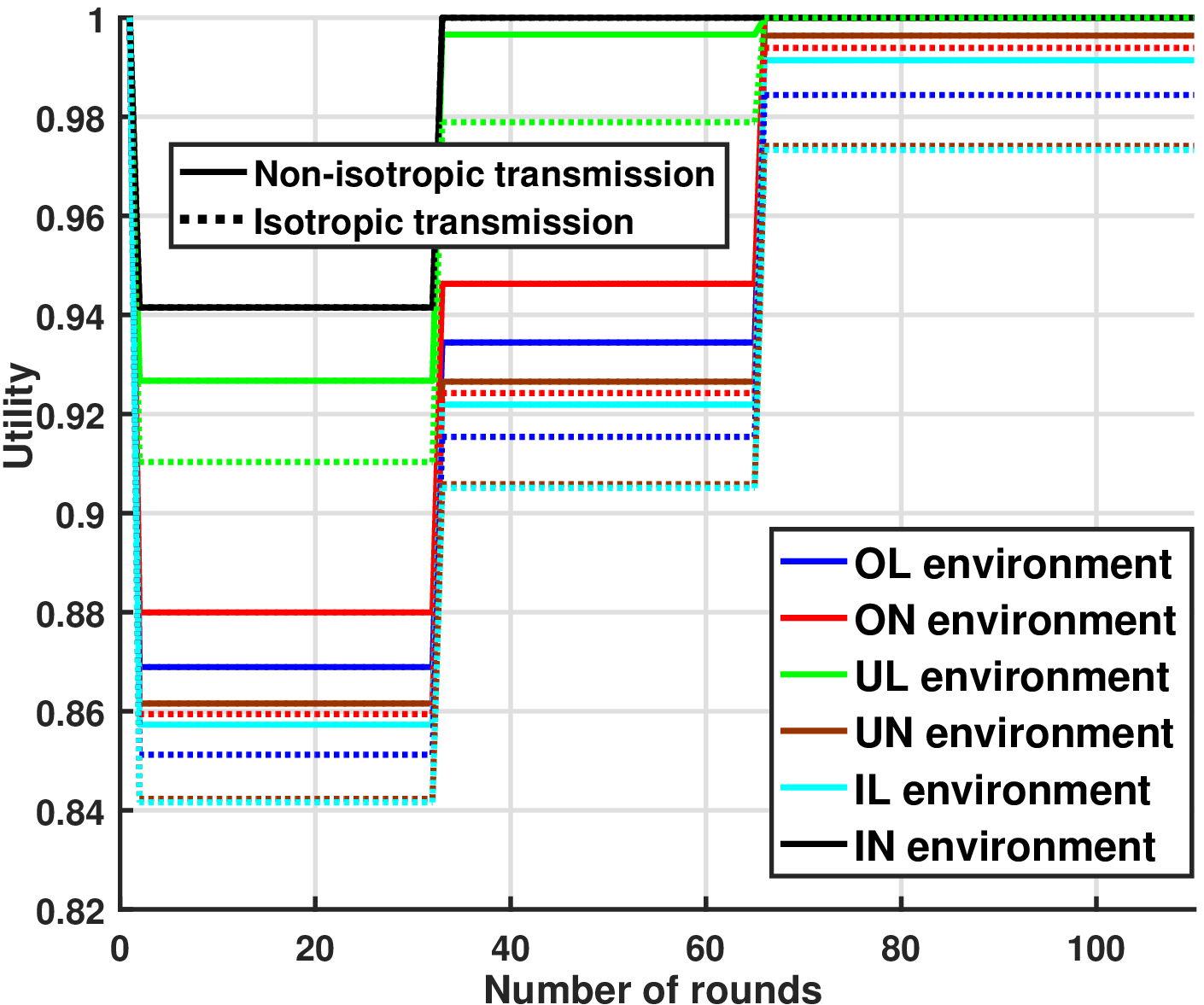}\vspace{-.05in} 
  \caption{DOI 1}
  \label{fig:1}
\end{subfigure}\hfil 
\begin{subfigure}{0.3\textwidth}
  \includegraphics[width=\linewidth]{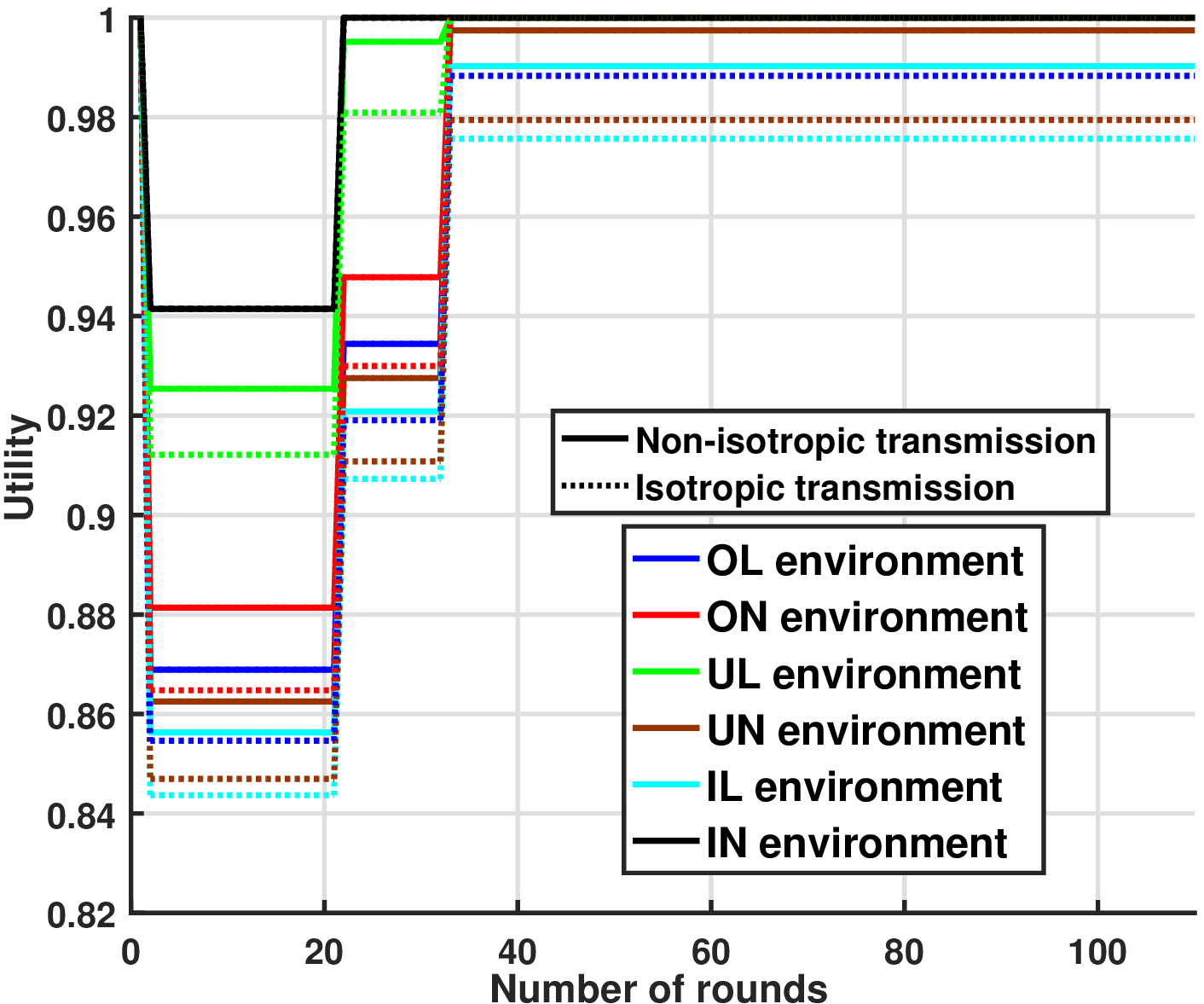}\vspace{-.05in}
  \caption{DOI 2}
  \label{fig:2}
\end{subfigure}\hfil 
\begin{subfigure}{0.3\textwidth}
  \includegraphics[width=\linewidth]{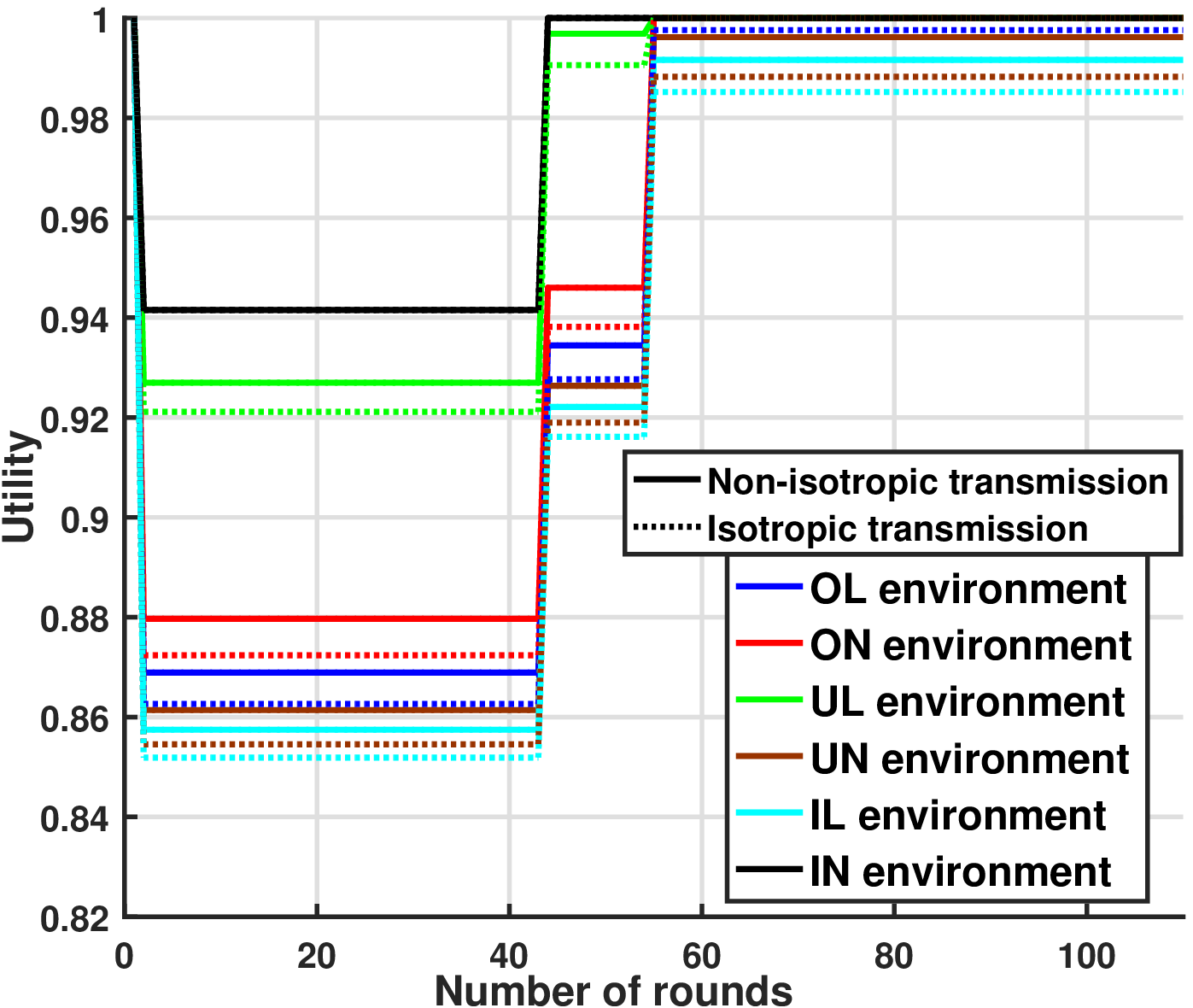}\vspace{-.05in} 
  \caption{DOI 3}
  \label{fig:3}
\end{subfigure}

\medskip
\begin{subfigure}{0.3\textwidth}
  \includegraphics[width=\linewidth]{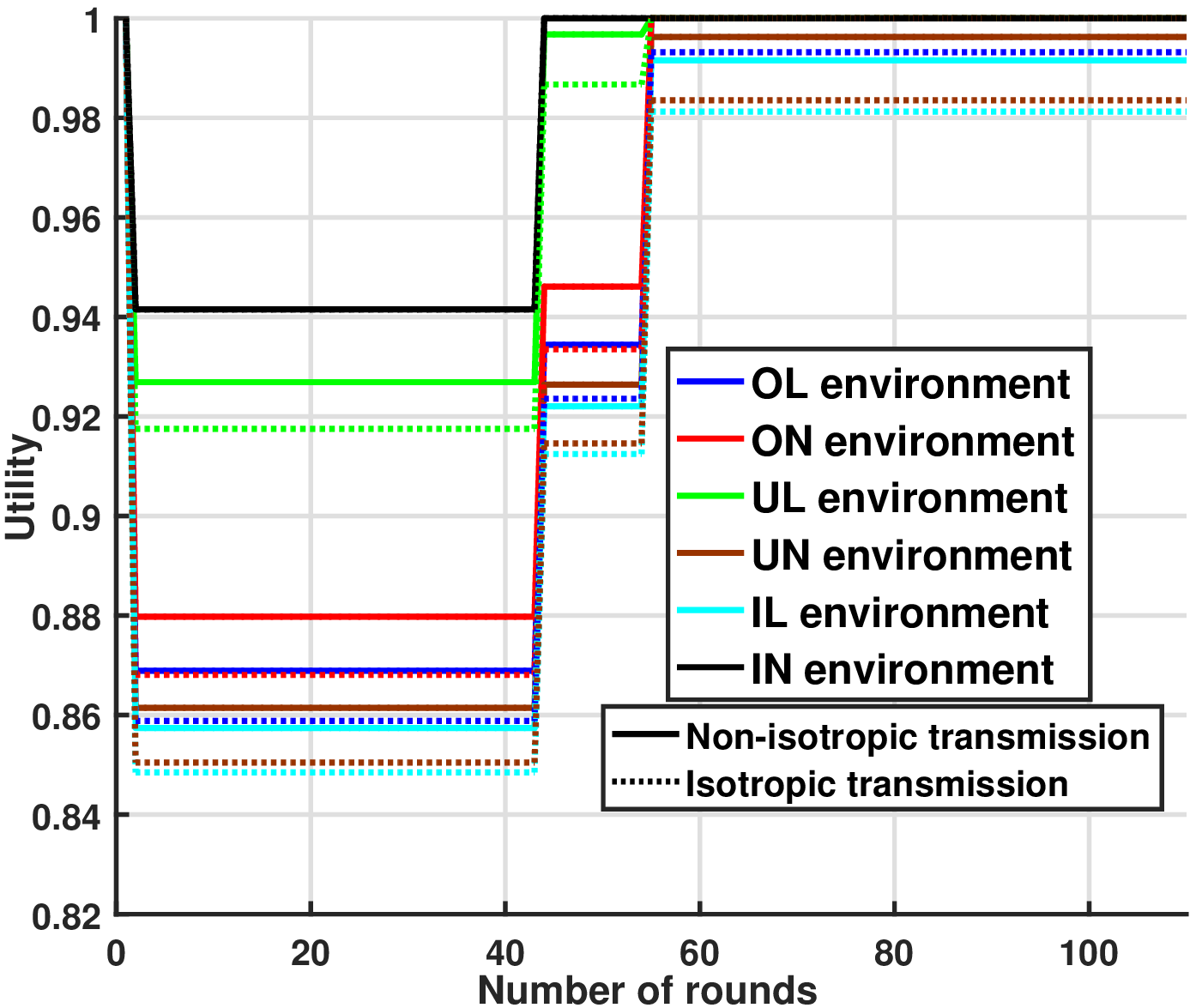}\vspace{-.05in} 
  \caption{DOI 4}
  \label{fig:4}
\end{subfigure}\hfil 
\begin{subfigure}{0.3\textwidth}
  \includegraphics[width=\linewidth]{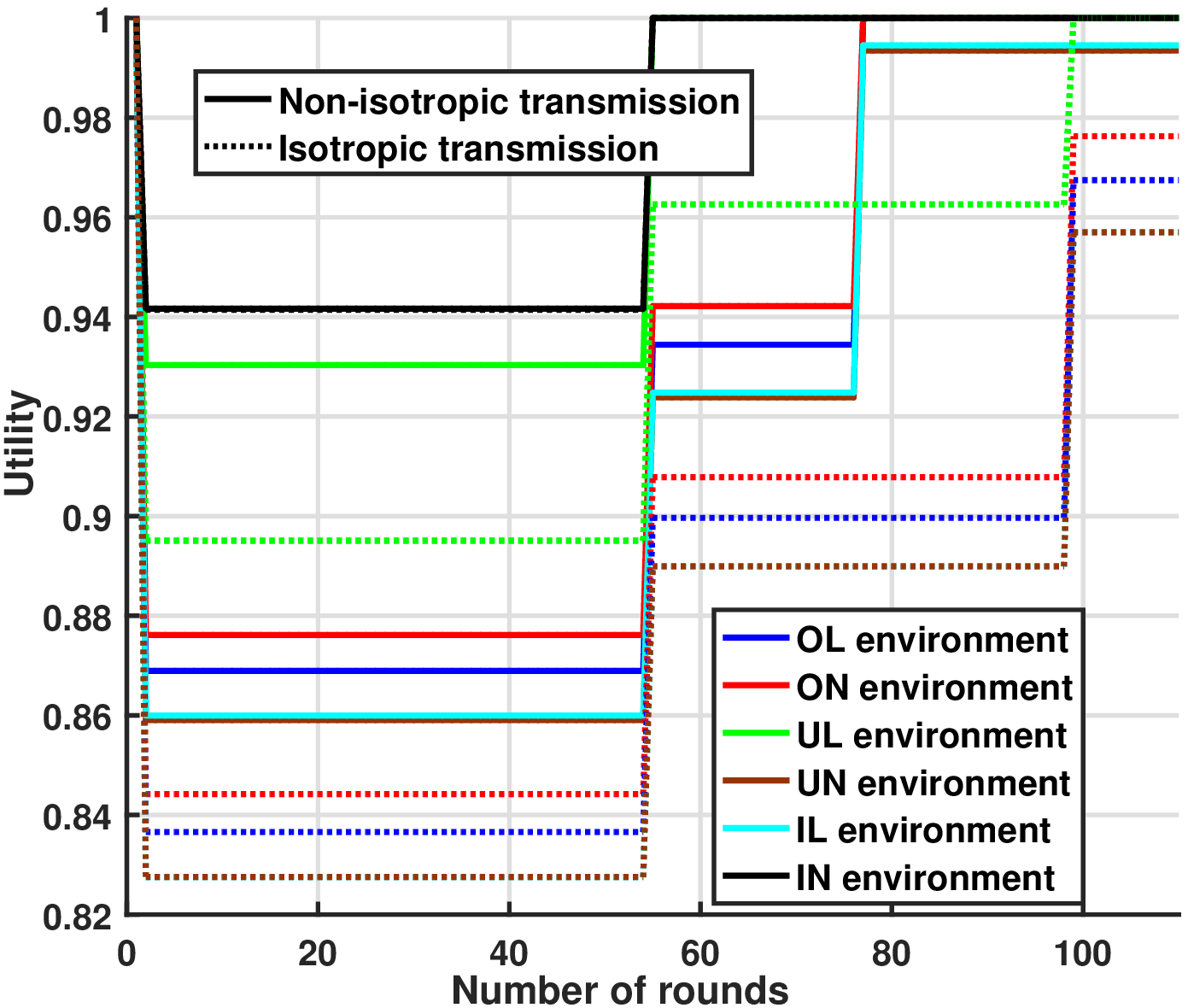}\vspace{-.05in} 
  \caption{DOI 4}
  \label{fig:5}
\end{subfigure}\hfil 
\begin{subfigure}{0.3\textwidth}
  \includegraphics[width=\linewidth]{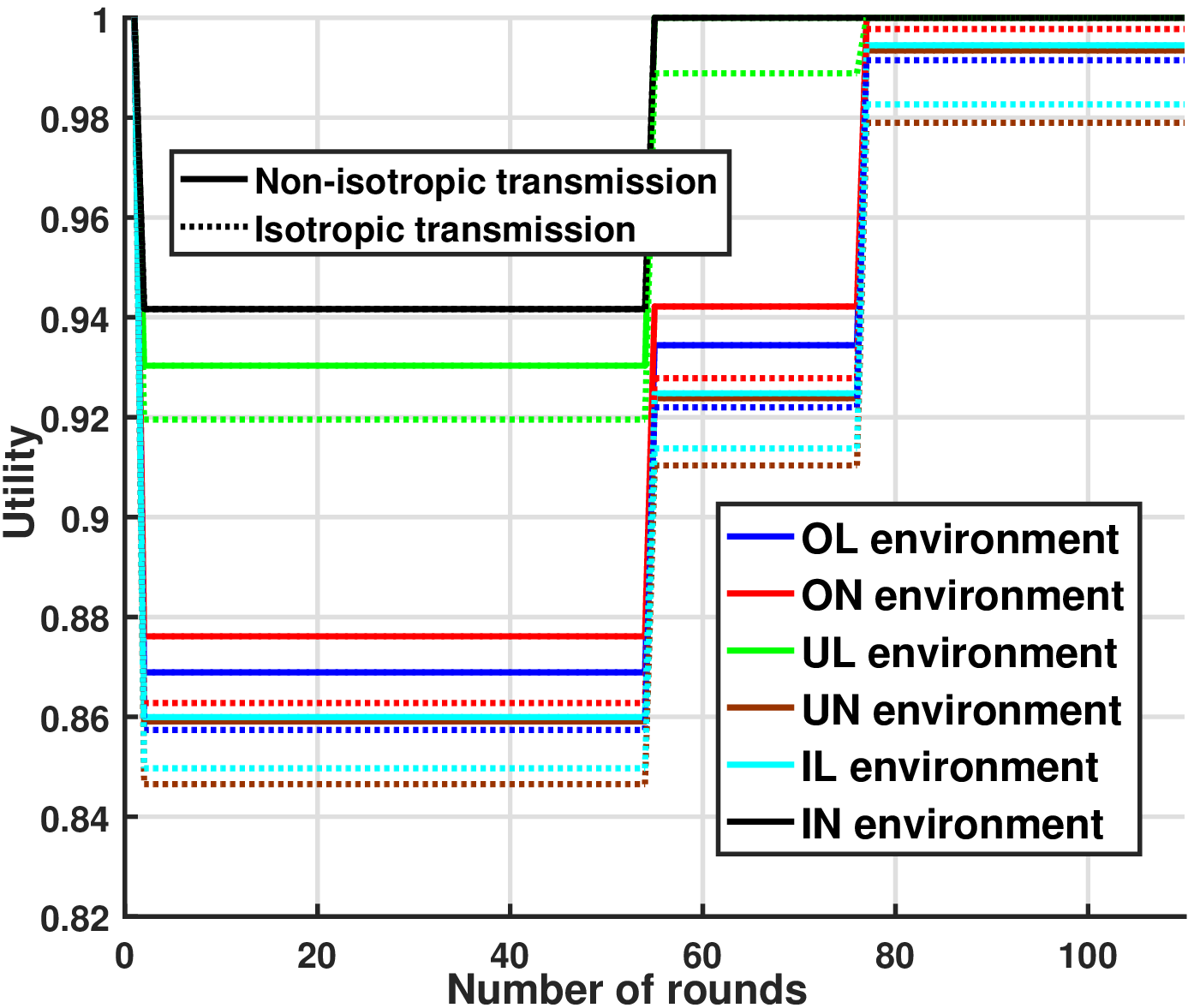}\vspace{-.05in} 
  \caption{DOI 6}
  \label{fig:6}
\end{subfigure}\vspace{-.1in}
\caption{Utility comparison of isotropic and non-isotropic packet transmission based on only the proposed repeated game over the six environments at every DOI, when number of malicious CMs equals 2.}
\label{DOI}
\end{figure*}

Figure \ref{DOI} shows a comparison between the obtained normalized utility functions of the isotropic and non-isotropic packet transmission using the proposed repeated game model over the six environments (OL, ON, UL, UN, IL, and IN) with each DOI measurements. Generally speaking, all the obtained results with any environment are close to their corresponding ones regardless of the DOI value. It is clear from Fig. \ref{DOI}a-d, with IN environment type, at which the values of $\sigma$ and $n$ are very close, the change of DOI does not affect the model performance as both the normalized utility of the isotropic and non-isotropic transmission are overlapped.  Interestingly, we can observe that the non-isotropic exceeds the isotropic packet transmission by only 1\% - 4\% that reflects the efficiency of the proposed model.

\begin{figure*}[!ht]
    \centering 
\begin{subfigure}{0.45\textwidth}
  \includegraphics[width=\linewidth]{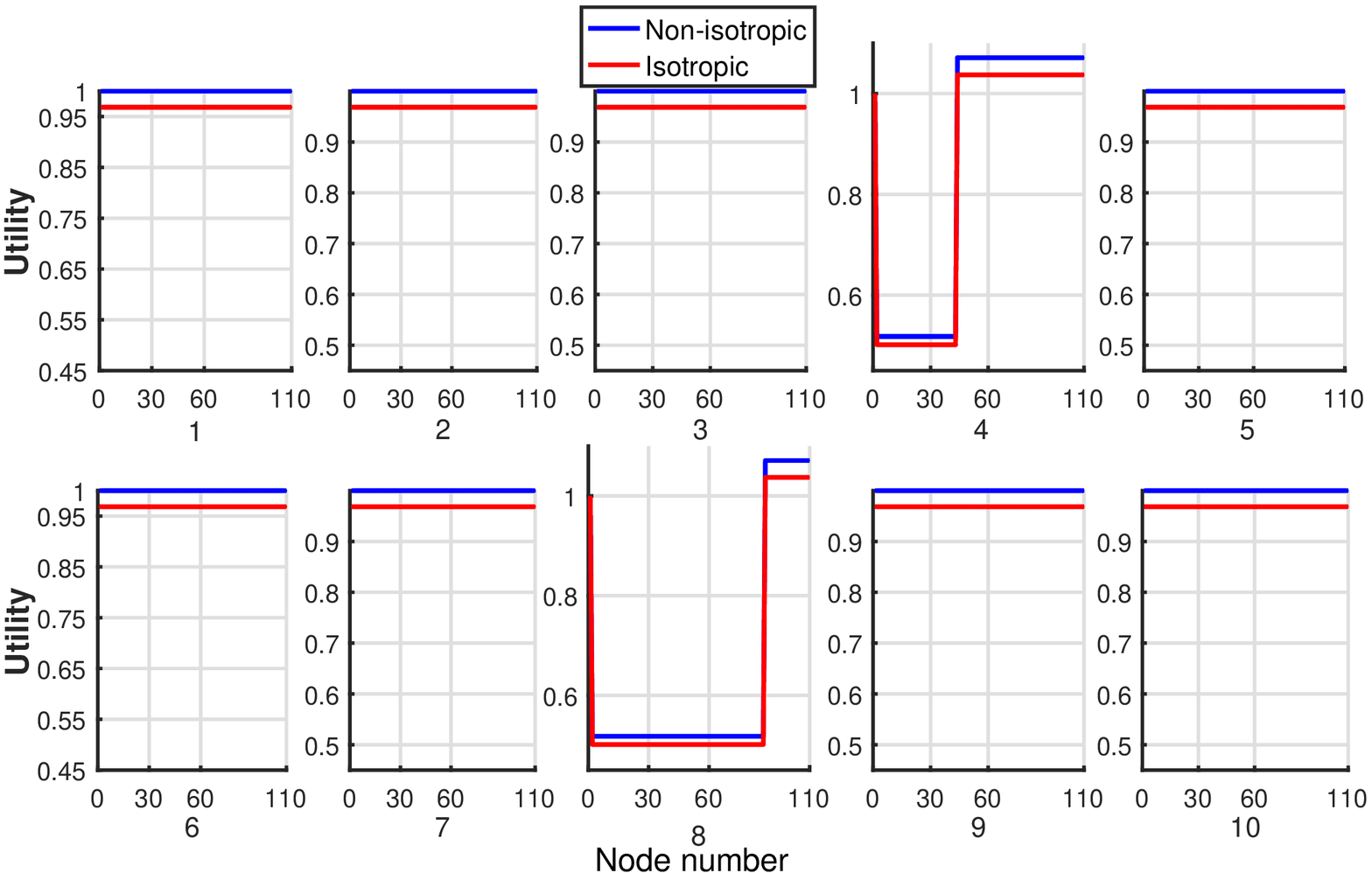} 
  \caption{UL}
  \label{fig:1}
\end{subfigure}\hfil 
\begin{subfigure}{0.45\textwidth}
  \includegraphics[width=\linewidth]{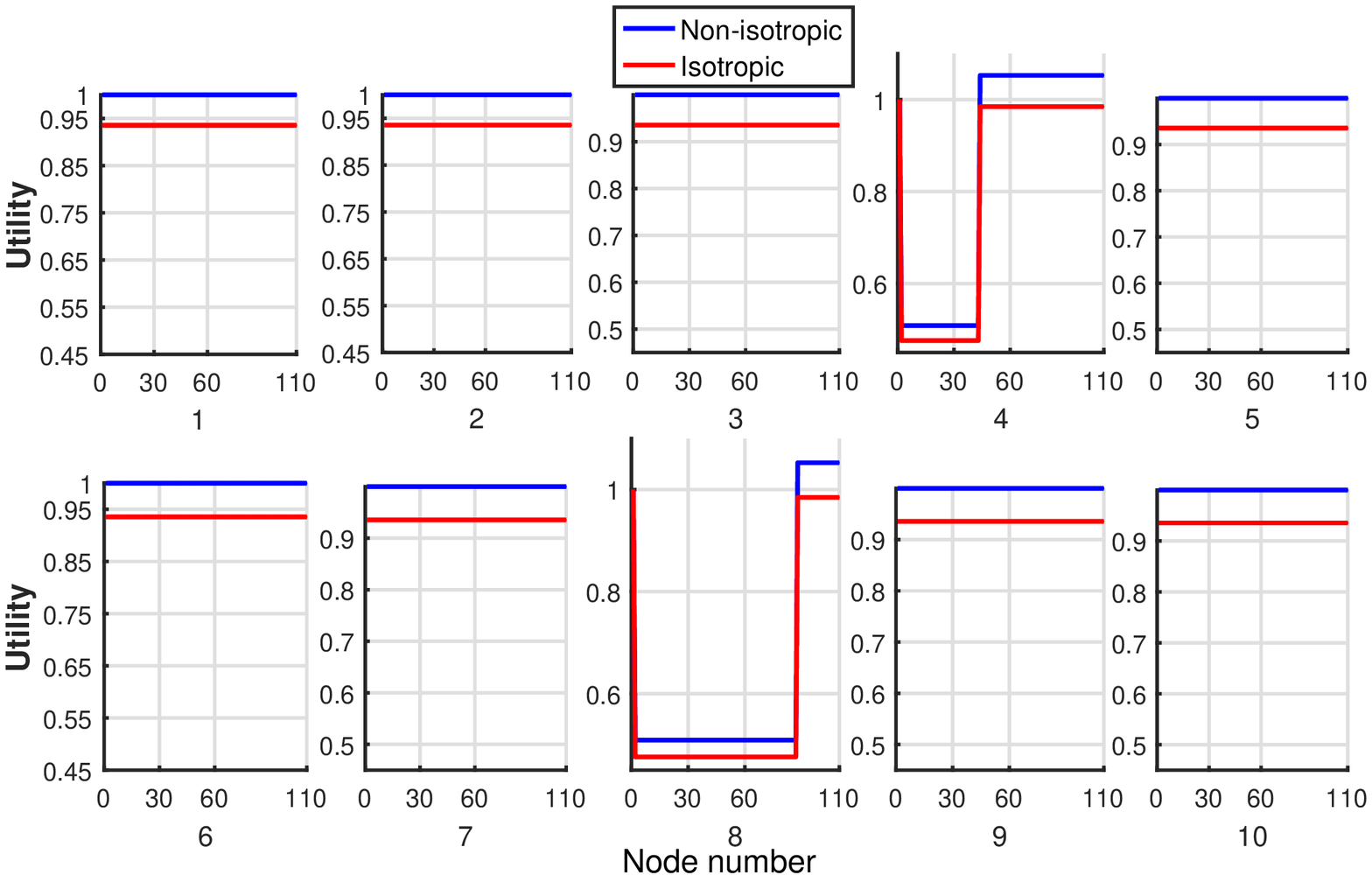} 
  \caption{ON}
  \label{fig:2}
\end{subfigure}
%
\caption{HW failure based on utility representation of isotropic and non-isotropic throughout (a) UL and (b) ON environments, 20\% of CMs suffer from HW failure.}
\label{HW_UL}
\end{figure*}

Detecting HW failure is discussed using both Fig. {\ref{HW_UL}}(a) and Fig. {\ref{HW_UL}}(b). At the equilibrium point, all the rational malicious attempts are refrained whether with isotropic or non-isotropic packet transmission. Afterward, any dropping packet or over packet transmission will be due to the HW failure not the SF attack.
Figure {\ref{HW_UL}}(a) shows the normalized utilities of the 10 CMs in which the values that exceed 100\% are due to the over packet transmission of the CMs that suffer from HW failure such as CM 4 and 8. The UL environment has the least path loss exponent that reflects the minimum effect on the isotropic radiation of packets. Therefore, the best-normalized utilities of using the proposed model with the isotropic packet transmission among the six environments are in the UL environment in which the utilities reach 96.8\%. Conversely, the ON environment represents the worse effect on the optimal isotropic normalized utilities, which reach only 93.5\%, as shown in Fig. {\ref{HW_UL}}(b).

\begin{figure}[h!]
\begin{center}
\includegraphics[width =.7\columnwidth]{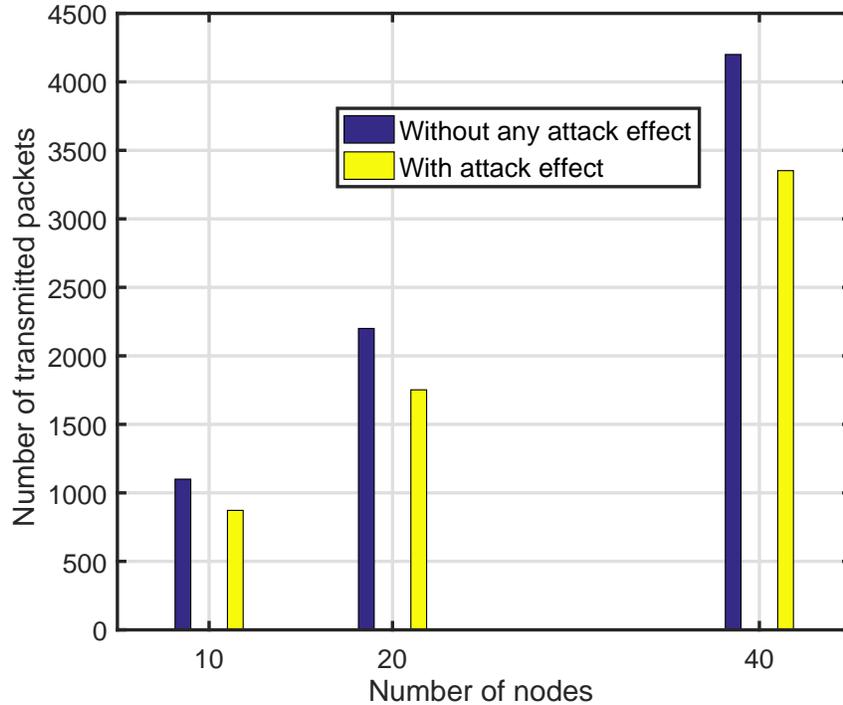}\vspace{-.1in}
\caption{Number of transmitted packets without and with attack presence using isotropic transmission over ON environment.}
\label{pkt}
\end{center}
\end{figure}

Afterward, the enhanced performance and model efficiency using the number of successfully transmitted packets have been depicted in Fig. \ref{pkt} using isotropic transmission over the ON environment representing the worst constraints. The number of transmitted packets counted at the equilibrium state based on the proposed repeated game is quite higher than the other three one-shot games as a result of the attack presence. In this regard, based on the proposed model along with the TDMA protocol, the CH can receive a much higher number of trusted packets from the CMs leading to the optimum DT in the six environments and DOIs. In other words, the model effectiveness has been achieved against the SF attack regardless of the environment and DOI types and whether the packet transmission is isotropic or non-isotropic. 

\begin{figure}[h!]
\begin{center}
\includegraphics[width =.7\columnwidth, height=7cm]{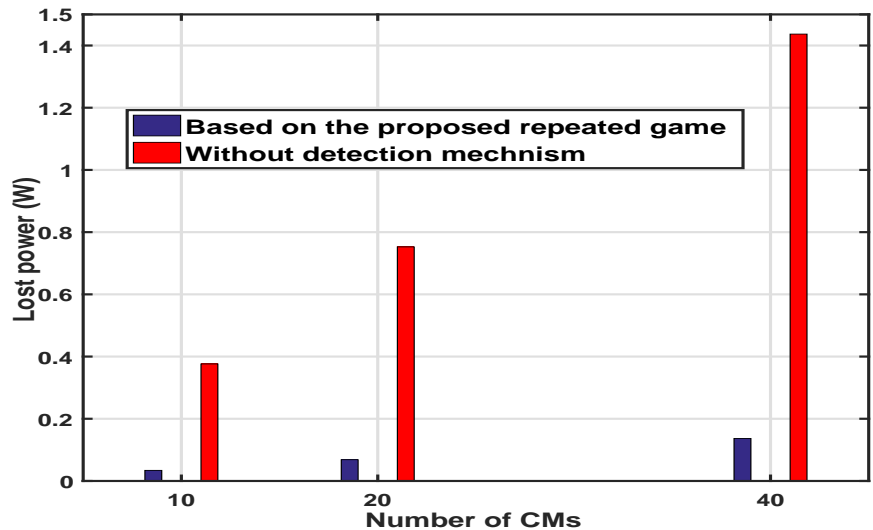}
\caption{Lost power for the selfish CMs due to the SF attack impact using isotropic transmission over ON environment.}
\label{power}
\end{center}
\end{figure}

Moreover, the comparison between the lost power using the proposed model and without detection mechanism against the SF attack is shown in Fig. \ref{power} over ON environment representing the worst constraints. If the proposed model has not been used, the network will be prone to lose a valuable amount of power due to the SF attack and the undiscovered CMs that suffer from HW failure, leading to battery power deterioration that increases with the increase of the number CMs as shown in Fig. \ref{power}. More particularly, the intensive power degradation is a result of giving negative acknowledgment by the CH to the malicious CMs and hence, they recursively transmit those packets that do not receive acknowledgments. Moreover, these CMs will not receive the Beacon to take a rest from the transmission or to go to the sleep mode. Therefore, the fraudulent behavior of the CMs will aggravatingly contribute to exhaust their battery lifetime.

\begin{figure}[h!]
\begin{center}
\includegraphics[width =.6\columnwidth, height=6.5cm]{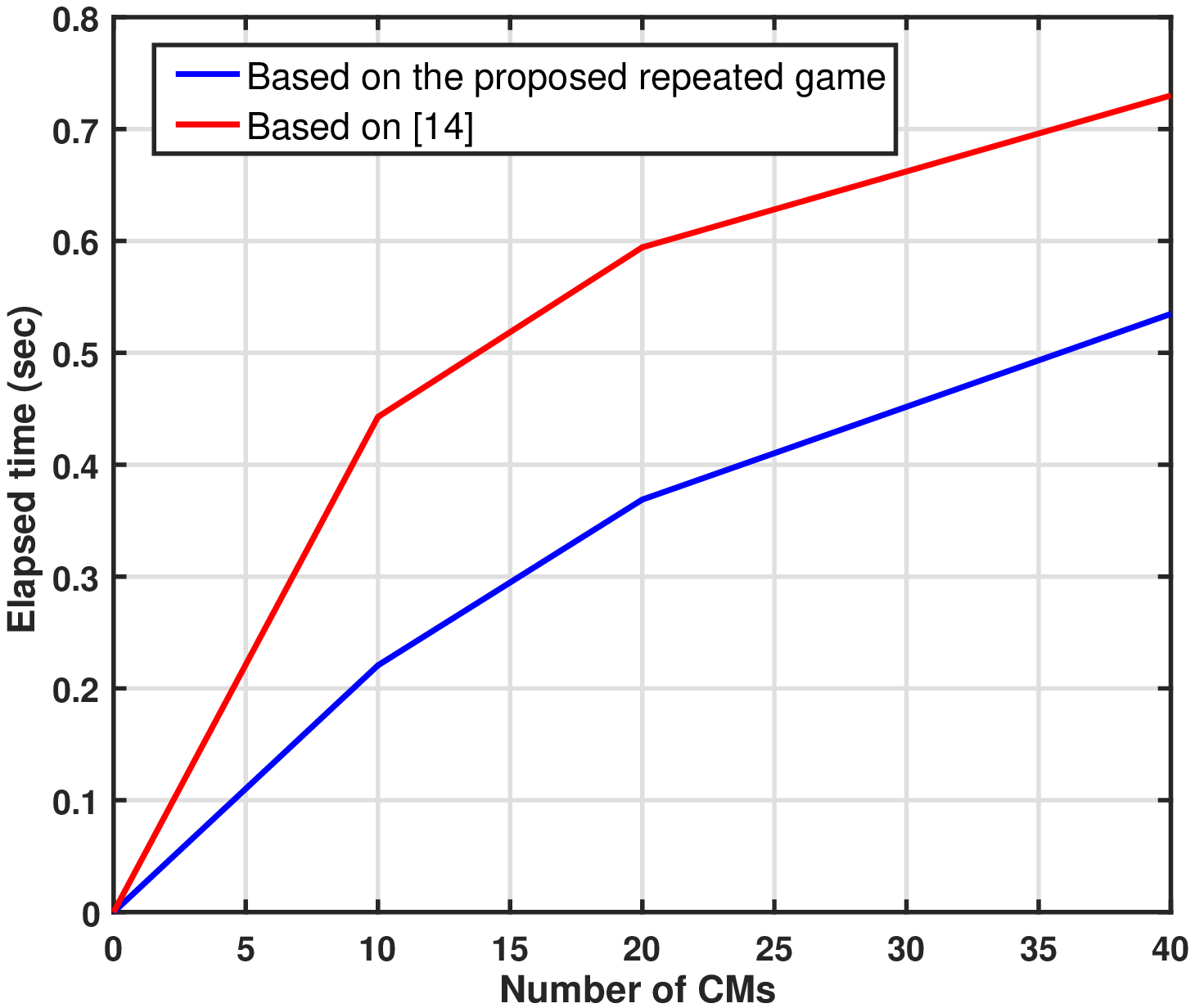}
\caption{Elapsed time comparison using isotropic transmission over ON environment.}
\label{elapsed_time}
\end{center}
\end{figure}

Figure {\ref{elapsed_time}} shows the relationship between the elapsed time and the number of CMs over the ON environment representing the worst constraints. The obtained results compare between the obtained results based on the proposed model and the corresponding work in \cite{abdalzaher2017using}. It is clearly shown that the time consumption using the proposed model is much less than the work in \cite{abdalzaher2017using} regardless of the number of CMs. Moreover, the proposed model presents an enhanced performance result regardless of increasing the number of CMs.

%


\section{Conclusion}\label{conclusion}
In this paper, we have proposed an efficient repeated game defense model along with the TDMA protocol to effectively detect the SF attack and HW failure in clustered WSNs-based IoT systems. The developed model can stimulate the malicious CMs that drop packets to react benevolently and hence, facilitate the HW failure detection to achieve the optimal DT. Consequently, the model solves a \textit{prisoner-dilemma} problem and is capable of attaining both the Pareto optimality and NE at the same state when the maximum DT exists. Moreover, the model proves beneficial in security enhancement, maximizing DT, and managing the consequently lost power leading to robust IoT networks. Simulation results demonstrate the effectiveness of the proposed model regardless of the environment, DOI, and isotropic and non-isotropic transmissions achieving robust WSNs-based IoT by preventing the packet drop due to the SF attack. Furthermore, the model can maximize the number of successfully transmitted packets. One of the future works is to evaluate our approach using testbed.  \vspace{-.05in}

\bibliographystyle{IEEEtran}
\bibliography{ref}

\end{document}